\documentstyle[preprint,eqsecnum,aps]{revtex}

\begin{document}
\draft


\title{\Large \bf Jet formation as a result of magnetic flux
                  tube--Kerr black hole interaction}
\author{Vladimir S. Semenov, Sergei A. Dyadechkin}
\address{Institute of Physics, State University, St. Petersburg
198504, Russia} %
\maketitle

\begin{abstract}                

Relativistic magnetohydrodynamics can be reformulated in terms of
magnetic flux tubes which turn out to obey equations of non-linear
strings. The string approach is applied to study how a test flux
tube falls into a Kerr black hole. Analytical treatment and
numerical simulations show that the leading portion of the falling
tube loses angular momentum and energy as the string brakes, and
to compensate for this loss, momentum and energy has to be
generated to conserve energy and momentum for the tube. Inside the
ergosphere the energy of the leading part can be negative, and the
rest of the tube then extracts energy from the hole. Increasing
centrifugal forces eject the part of the tube with extra positive
energy  from the ergosphere after a time producing a relativistic
jet.

\end{abstract}

\narrowtext

\section{ INTRODUCTION}

The mechanism by which energy is extracted from a black hole and
astrophysical jet forms is a key problem in our understanding of
phenomena such as active galactic nuclear, quasars, and probably
gamma ray bursts.  The present paper considers an alternative
approach to the famous Blanford-Znajek process \cite{1}. This is
described in \cite{2,3}, which is an extension of the Penrose
mechanism \cite{4} for extracting energy from a rotating black
hole by a relativistic string. It was shown that the falling
magnetic flux tube can effectively generate negative energy, but
the parts of the tube with both negative and compensated positive
energy are localized inside the ergosphere and almost does not
show up outside. The aim of this paper is to present examples of
how a falling flux tube can also produce a relativistic jet.

\section{ FORMULATION OF  RMHD IN TERMS OF STRINGS }
\label{SR}

  An appropriate formulation for studying the behaviour of
flux tubes can be achieved through the introduction of Lagrangian
coordinates \cite{2,3} into the following relativistic
magnetohydrodynamic (RMHD) equations \cite{5}
\begin{eqnarray}
&\nabla_i\rho u^i=0, \label{1}
\\
& \nabla_i T^{ik}=0, \label{2}
 \\
&\nabla_i(h^i u^k-h^k u^i)=0. \label{3}
\end{eqnarray}
Here  $u^i$ is the time-like vector of the 4-velocity, $u^i
u_i=1$,
 $h^i=\ast F^{ik}u_k$ is the space-like 4-vector of the magnetic field,
$h^i h_i<0$, $\ast F^{ik}$ is the dual tensor of the
electromagnetic field, and $T^{ij}$ is the energy-momentum tensor:
\begin{eqnarray}
&T^{ij}=Q u^i u^j - P g^{ij}-\frac {1}{4\pi} h^i h^j, \label{4}
\end{eqnarray}
where
\begin{eqnarray}
P \equiv p-\frac {1}{8\pi}h^k h_k,\,\ Q \equiv p+\varepsilon-\frac
{1}{4\pi}h^k h_k. \label{5}
\end{eqnarray}
Here $p$ is the plasma pressure, $P$ is the total (plasma plus
magnetic) pressure, $\varepsilon $ is the internal energy, and
$g_{ik}$ is the metric tensor with signature $(1,-1,-1,-1)$.

We can find a function $q$ such that $\nabla_i q h^i = 0$, then
using (\ref{1}) the Maxwell equation (\ref{3}) can be rewritten in
the form of a Lie derivative
\begin{equation}
\frac{h^i}{\rho}\nabla_i \frac{u^k}{q}=
\frac{u^i}{q}\nabla_i \frac{h^k}{\rho},\label{6}
\end{equation}
and we can therefore introduce Lagrangian coordinates $\tau,
\alpha$ such that \cite{6}
\begin{eqnarray}
x^i_\tau \equiv \frac{\partial {x^i}}{\partial \tau} =
\frac{u^i}{q}, \,\ x^i_\alpha \equiv \frac{\partial
{x^i}}{\partial \alpha} = \frac{h^i}{\rho}
 \label{7}
\end{eqnarray}
with new coordinate vectors $u^i/q, h^i/\rho$ tracing the
trajectory of a fluid element and the magnetic field in a flux
tube. Since the introduction of these coordinates relies on the
frozen-in property of the plasma, we will refer to them as
frozen-in coordinates \cite{2,3}. The mass coordinate $\alpha$
along the relativistic flux tube has the sense of a mass of the
plasma for a tube with unit flux in the proper system of
reference. The second coordinate $\tau$, the string time, is not
any more Lagrangian or proper time, but is just a time-like
parameter which traces the flux tube in the space-time of general
relativity. The functions $x^i(\tau, \alpha)$ sweep the 2D
worldsheet in the space-time continuum which consists of
trajectories of the fluid elements for $\alpha=\mbox{const}$, and
the magnetic field lines for $\tau=\mbox{const}$.

Since the following relations are valid in the new coordinates
$\tau, \alpha$ \cite{3}:
\begin{equation}
\frac{u^i}{q}\nabla_i= \frac{\partial}{\partial{\tau}},\quad
\frac{h^i}{\rho}\nabla_i= \frac{\partial}{\partial{\alpha}}, \label{rel3c}
\end{equation}
 the energy-momentum equation
(\ref{2}) can be rearranged to form a set of string equations:
\begin{eqnarray}
& &-\frac{\partial }{\partial \tau}
 \left(\frac{Q q}{\rho}x^l_\tau\right)-
 \frac{Q q}{\rho}\Gamma^l_{ik}x^i_\tau x^k_\tau  \nonumber \\
& & \hspace*{1cm} + \frac{\partial }{\partial \alpha}
 \left(\frac{\rho}{4 \pi q}x^l_\alpha\right)+
 \frac{\rho}{4\pi q}\Gamma^l_{ik}x^i_\alpha x^k_\alpha=
 -\frac{g^{il}}{\rho q} \frac{\partial P}{\partial {x^i}}, \label{9}
\end{eqnarray}
where $\Gamma^l_{ik}$ is the Christoffel symbol. Generally
speaking, the total pressure $P({x^i})$ is unknown in advance, but
fortunately there are a set of problems when $P({x^i})$ can be
considered as a given function. In this case a test flux tube may
represent MHD flow as a whole. It is  known that characteristics
of the string equations (\ref{9}) are Alfv\'en and slow waves
\cite{3}. Hence, the fast wave which is also characteristic of the
general MHD system of equations (\ref{1}-\ref{3}) \cite{5}, is
left out in the string approach. The physical reason for this is
clear: the fast wave is produced by gradients in the total
pressure, but $P({x^i})$ is fixed in equations (\ref{9}), and
there is no driving mechanism for fast waves.

Therefore, MHD problems which involve significant variations of
total pressure such as gasdynamic-type explosion, are unlikely to
be applicable using the string equations (\ref{9}). But processes
of accumulation and relaxation of Maxwellian tensions can very
often be described by the string approach since the total pressure
does not seem to vary appreciably, like in Alfv\'en wave, for
example. Hence, $P(x^i)$ may be considered as a given function or
at least may be determined by a perturbation method. As a result,
the general 4D RMHD problems can be reduced to investigation of
the behaviour of a 2D test flux tube/string. Such a method has
been successfully applied to problems in magnetospheric \cite{7},
solar \cite{8} and astrophysical \cite{3} plasmas.

\section {Flux Tube --  Kerr Black Hole Coupling}

The Kerr metric is given in Boyer-Lindquist coordinates by
\cite{6}
\begin{eqnarray}
&ds^2=(1-\frac{2 M r}{\Sigma})dt^2-
\frac{\Sigma}{\Delta}dr^2-\Sigma d\theta^2-
 \nonumber\\
&(r^2+a^2+\frac{2 M r a^2}{\Sigma}\sin^2\theta)\sin^2\theta d\varphi^2+
\frac{4 M r a}{\Sigma} \sin^2\theta d\varphi dt, \label{15}
\end{eqnarray}
where
\begin{eqnarray}
\Delta=r^2-2 M r +a^2, \,\ \Sigma=r^2+a^2\cos^2\theta. \label{16}
\end{eqnarray}
Here $M$ and $a$ are the mass and specific angular momentum of the
hole, respectively, and the units are chosen such that $c=1, \ \
G=1$.

We will now consider a test magnetic flux tube falling into a Kerr
black hole (Figure 1a). We solve the string equations (\ref{9})
numerically using TVD scheme similar to \cite{3}. We normalize
plasma density to its initial value $\rho_0$, length to the radius
of the static limit surface in the equatorial plane $r_g$,
velocity to the light speed $c$, time scale to $r_g/c$, magnetic
field to $\sqrt{4\pi \rho_0}c$, the plasma pressure and the energy
density to $\rho_0 c^2$, $\alpha$ to  $r_g \sqrt{\rho_0/(4\pi
c^2)}$. The number of grid points along the string $-20<\alpha<20$
is chosen to be 4000. Only the central part $-10<\alpha<10$ will
be shown in Figures. The black hole is supposed to be near the
extreme rotation $a=.99 M$.

We need to specify the total pressure in the string approach. The
dependence $P(r)={c_P}/{(r-r_h)^2}$ has been used in \cite{3},
where $c_P$ is a constant. In this case both negative and the most
of positive energy was confined inside the ergosphere. Here we
increase the pressure gradient $P(r)={c_P}/{(r-r_h)^3}$ and as a
result are able to observe ejecta from the hole.

It is convenient to chose an initial flux tube outside the
ergosphere so that each element of the tube has no angular
momentum. Hence, if there is no magnetic field, no tube element
would interact with any other element, and as it falls into the
black hole it would rotate with an angular velocity
$\omega_0=-g_{t\varphi}/g_{\varphi\varphi}$ of fiducial observers
\cite{6}.

Because of inhomogeneous rotation of the Kerr geometry, the flux
tube becomes stretched and twisted (Figure 1a), and gravitational
energy is partially converted into magnetic energy of the swirling
tube. The strong magnetic field slows down rotation of the leading
part of the flux tube as it falls, hence this part of the tube
obtains negative momentum (Figure 2a). Angular momentum of the
tube as a whole needs to be conserved, and to compensate this a
corresponding positive angular momentum is generated (Figure 2a).

Simulations show that the stretching and twisting of the falling
flux tube is most pronounced  inside the ergosphere (Figure 1b)
where $g_{tt}<0$ and the energy of the central part of the tube
with negative momentum turns out to be also negative (Figure 2b).
Since the energy of the tube as a whole has to be conserved, the
part of the tube with positive energy gains  an energy greater
than the initial tube energy (Figure 3). This is a variant of the
Penrose process \cite{4}, but now we do not need to invoke the
decay of particles, since just a single tube can extract energy
from the hole.

The process of redistribution of angular momentum is continuous:
the deeper the tube falls in the hole, the more the tube is
stretched, the stronger the magnetic field gets, the slower the
central part of the tube rotates, the more negative momentum is
generated, and the more positive momentum is created (Figure 2b).
It turns out that both parts with negative and most of the
positive energy are localized at this first stage in the narrow
layer near the event horizon \cite{3}. Note that the part of the
flux tube with positive momentum and with extra positive energy
has to spin faster and faster. Eventually,  increasing centrifugal
forces eject plasma from the ergosphere producing relativistic jet
with the Lorentz factor $\Gamma\sim3$ (Figures 1c, 2c). Note that
the whole string remains uninterrupted and simple connected.

When the part of the tube with positive energy is ejected, the
remaining part with negative energy can not leave the ergosphere.
It continues  spinning around the event horizon producing the
spiral magnetic field and extracting rotational energy of the
hole. During this quasistationary stage the rate of energy
extraction is approximately equal to the negative energy creation
rate and nearly twice as big as energy accretion rate (Figure 3).

\section{ Conclusion}
\label{Su}

After submission of our paper we got to know the results of
general RMHD numerical simulations modeled accretion of a plasma
into a Kerr black hole \cite{9}. The comparison of the results
obtained with two different methods show that they are in a good
agreement. This includes the stretching and twisting of magnetic
flux tubes around the event horizon, generation of negative
angular momentum and negative energy of the plasma near the event
horizon, appearance  of compensated positive momentum and positive
energy. The problem under consideration is a stiff one, MHD
parameters increase very fast inside the ergosphere,  and at some
stage errors of calculation started to grow exponentially.
Therefore the authors of \cite{9} were able to model only early
stage of plasma accretion into a rotating black hole and they had
to stop calculations at $t=7 r_g/c$. At this stage which roughly
correspond to our Figure 1b, the plasma jet is not yet formed,
instead they observed MHD wave which transferred  electromagnetic
energy outside the ergosphere. We also found this effect in our
previous study \cite{3}, but unfortunately it turns out that MHD
wave extract energy at relatively low rate compared to the
negative energy creation rate. The fact is that the infall
velocity of the plasma is approximately equal to the speed of the
MHD wave.

A relativistic flux tube (i.e. string) is more simple object as
the whole MHD flow. Thus, we were able to continue our
calculations to later time $\tau=20 r_g/c$ and to observe a real
plasma ejecta from the hole. We also had a similar problem with
the code and we needed to enhance all factors stimulating plasma
ejection from the ergosphere such as the angular momentum of the
hole and the total pressure gradient. However, the Kerr black hole
is so powerful object that for another distribution of the total
pressure the jet seems to appear just a bit later anyway.
Unfortunately we can not simulate this more realistic situation so
far due to the code problem.

The physics of the energy extraction from a rotating black hole
and formation of a cosmic jet is rather simple. In the spinning
Kerr geometry the leading part of the falling flux tube
progressively loses angular momentum and energy as the string/tube
brakes, which leads to creation of negative energy inside the
ergosphere. To conserve energy and angular momentum for the tube
as a whole, the positive energy and angular momentum has to be
generated for the trailing part of the tube. The MHD wave can not
effectively remove generated positive energy from the ergosphere,
and as a result the part of the tube with positive angular
momentum has to rotate faster and faster in the course of time.
Eventually the growing centrifugal forces push this part of the
tube from the ergosphere producing plasma ejecta and extracting
spin energy from the hole.

It is worth  discussing the string model in relation to the
Blandford-Znajek mechanism \cite{1}. The string process is
inherently time-dependent because it is based on differential
rotation even for the quasi-stationary regime. It is well known
that for a purely steady-state axisymmetric configuration, the
flux surface must have rigid rotation \cite{10} which is not the
case for the string mechanism. There is only a tendency to rigid
rotation, which is clear from the redistribution of the angular
momentum, i.e., the leading part of the falling tube spins slower,
the trailing part faster. But to establish exactly rigid rotation
takes an infinite time, which can hardly happen in reality.
Therefore, the string mechanism differs from the Blandford-Znajek
one based on steady-state and axisymmetrical patterns. Besides,
the Blandford-Znajek process needs a magnetic field embedded in
the hole's event horizon while the flux tube in the string
approach does not reach the horizon at all in our simulations,
which emphasizes the difference even more. However, the main idea
of Blandford and Znajek  \cite{1} considering the magnetic forces
as the most appropriate to couple the black hole's spin to
external matter is also valid for the string approach.

There is a direct extension of the theory described above to
cosmic strings. Formally the equations of motion for ordinary
cosmic strings generated by the Nambu-Goto action \cite{11}, and
especially for superconducting cosmic strings, are the same as
those for the flux tubes. Therefore, we can expect extraction of
spin energy of the hole by a cosmic string, as it was first
pointed out in \cite{12}.

{\sl Acknowledgements}: We are thankful to Referees for useful
criticism and reference \cite{9}, and to I. V. Kubyshkin and R. P.
Rijnbeek for helpful discussions. This work is supported by the
Russian Foundation of Basic Research, grant No.
\mbox{01-05-64954}, by INTAS-ESA, grant No. 99-01277. SAD was
suported by INTAS, grant No. YSF-80.

\newpage

\setcounter{figure}{0} 

\begin{figure}
\caption{Time evolution of the falling flux tube is shown for the
initial position of the tube (dashed line) and the moment of
encounter of the static limit surface (solid line)(a),for the
beginning of the ejecta (b), and for the final time of simulation
(c). Event horizon is depicted in the center of the pictures. }
 \label{fig:1}
\end{figure}

\begin{figure}
\caption{Numerical results for a flux tube falling into a Kerr
black hole  as functions of the mass parameter $\alpha$.
Distributions of the density of the angular momentum (a), of the
density of the energy (b), and of the radial Boyer-Lindquist
coordinate $r$ along the string are shown for the following string
times: $\tau_1$ is the initial time, $\tau_2$ is the moment of
encounter of the static limit, $\tau_3$ is the beginning of the
negative energy creation, $\tau_4$ is the beginning of the ejecta,
and $\tau_5$ is the final time of simulation.}
 \label{fig:2}
\end{figure}

\begin{figure}
\caption{Behaviour of the positive (1), total (3), and negative
(5) energy of the string in the course of string time. The curves
(2) and (4) shows evolution of the energy of the part of the tube
outside and inside the ergosphere, respectively.}
 \label{fig:3}
\end{figure}

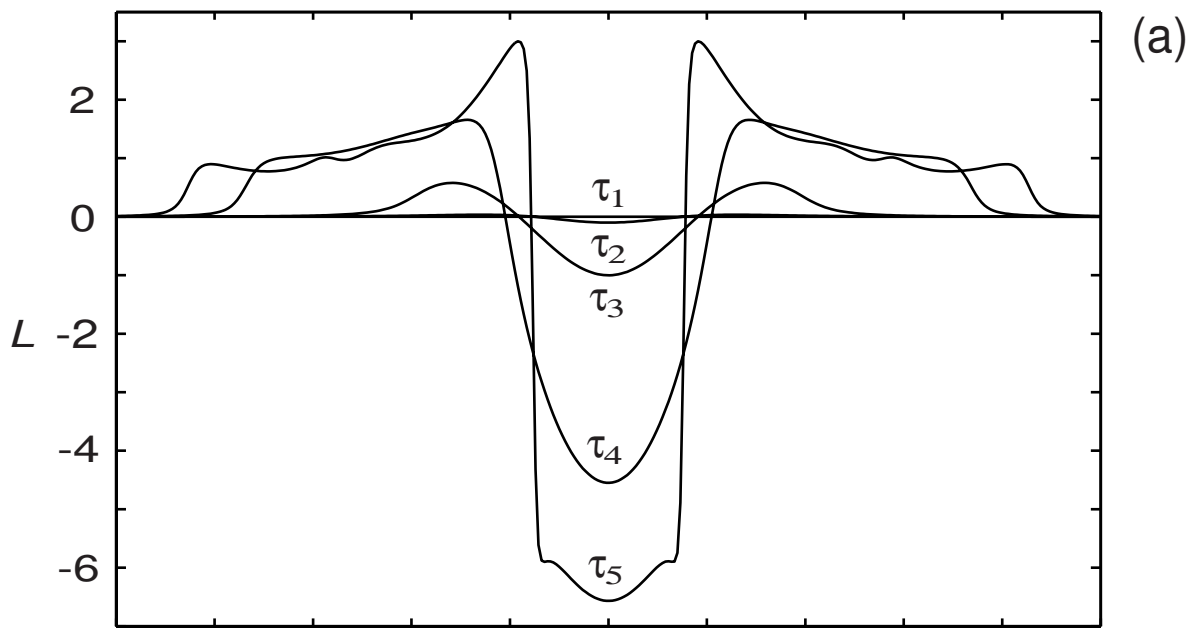
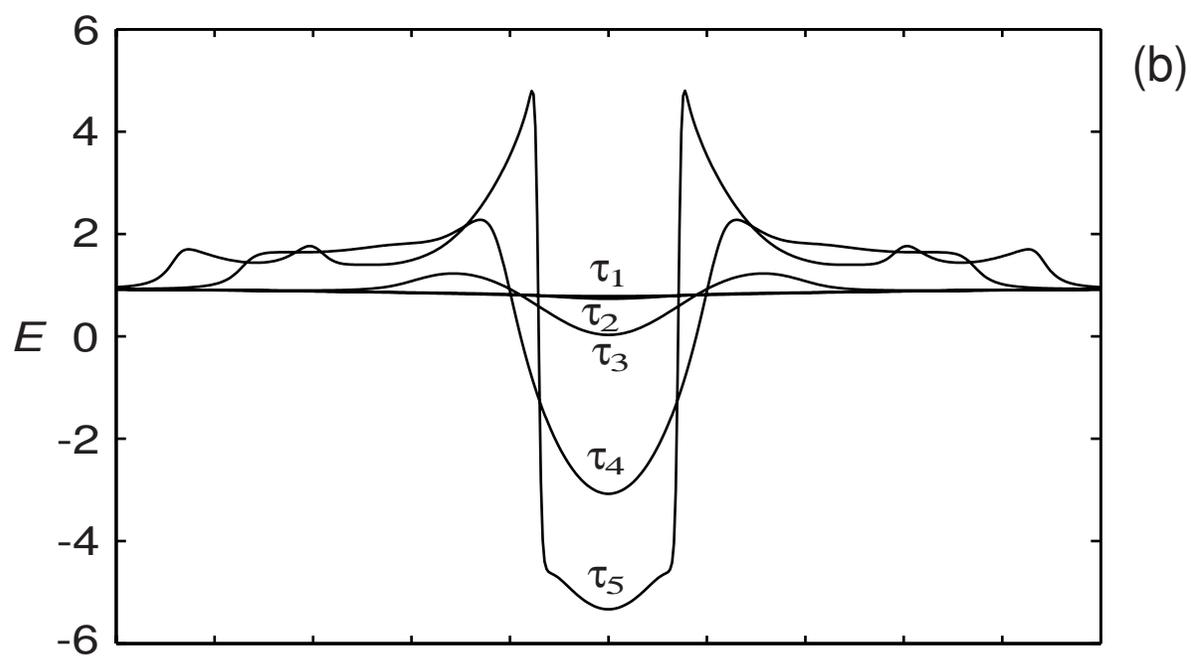
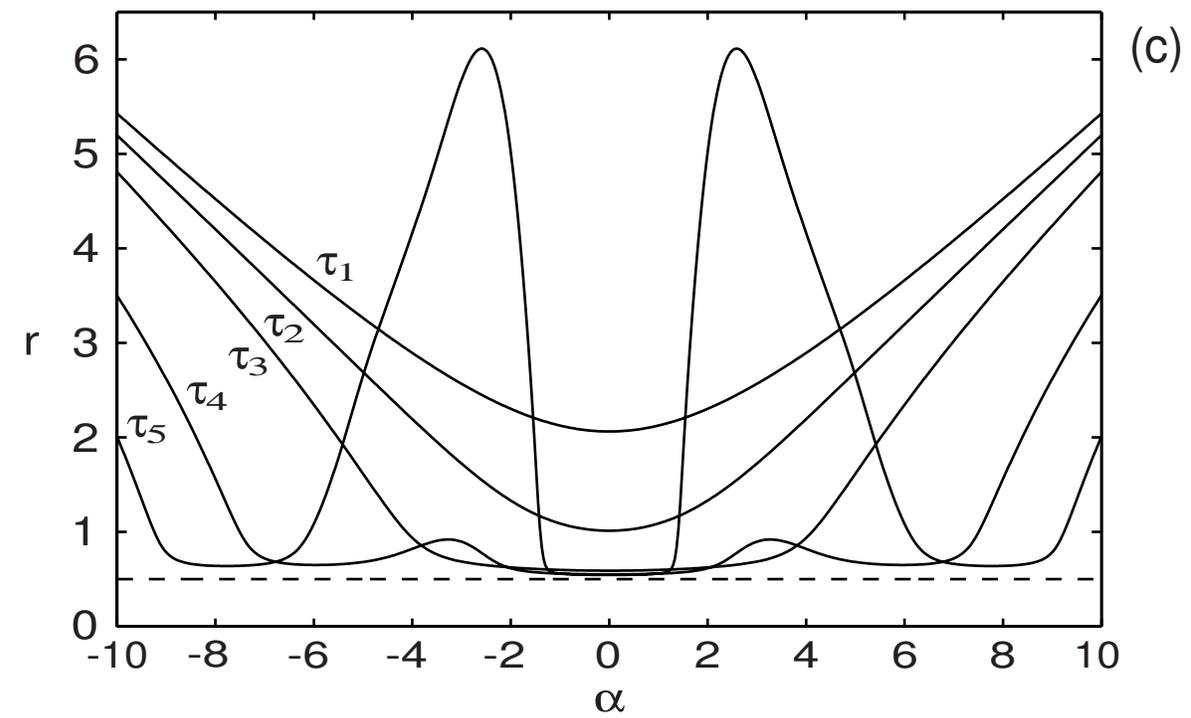
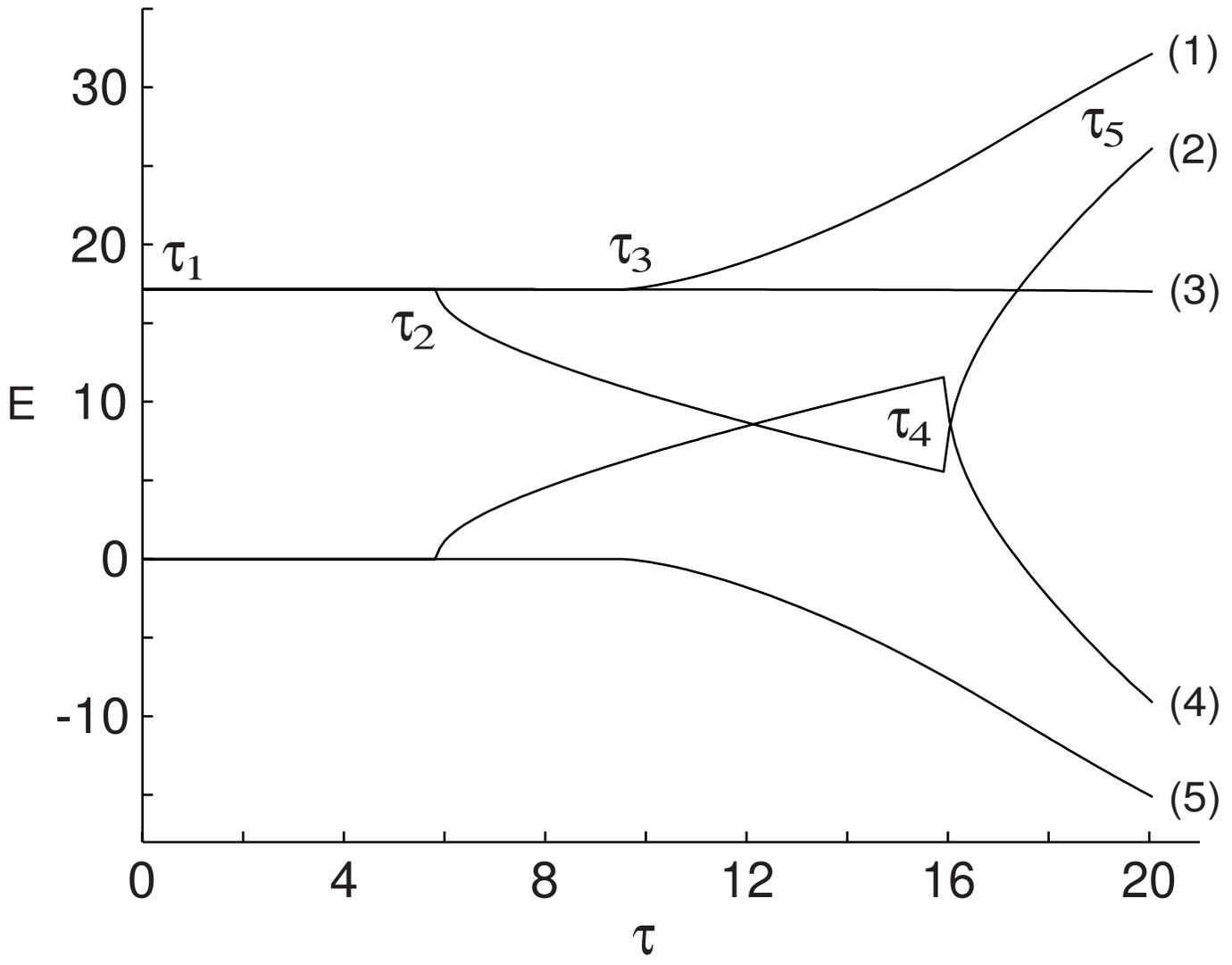
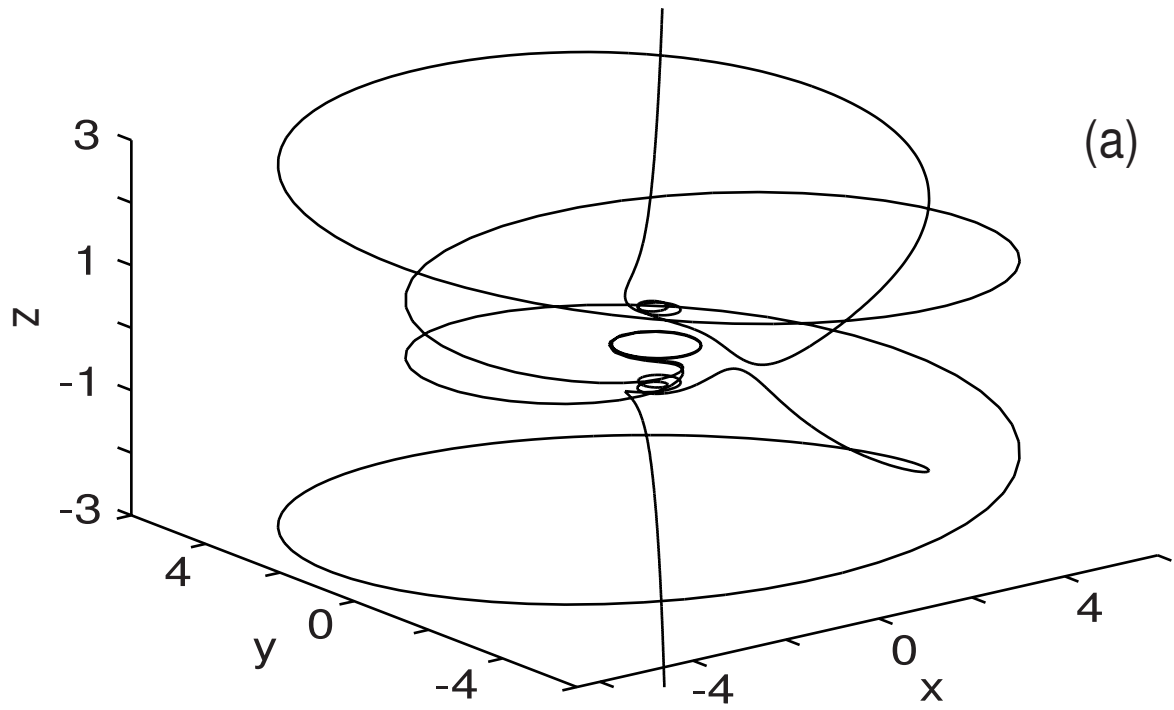
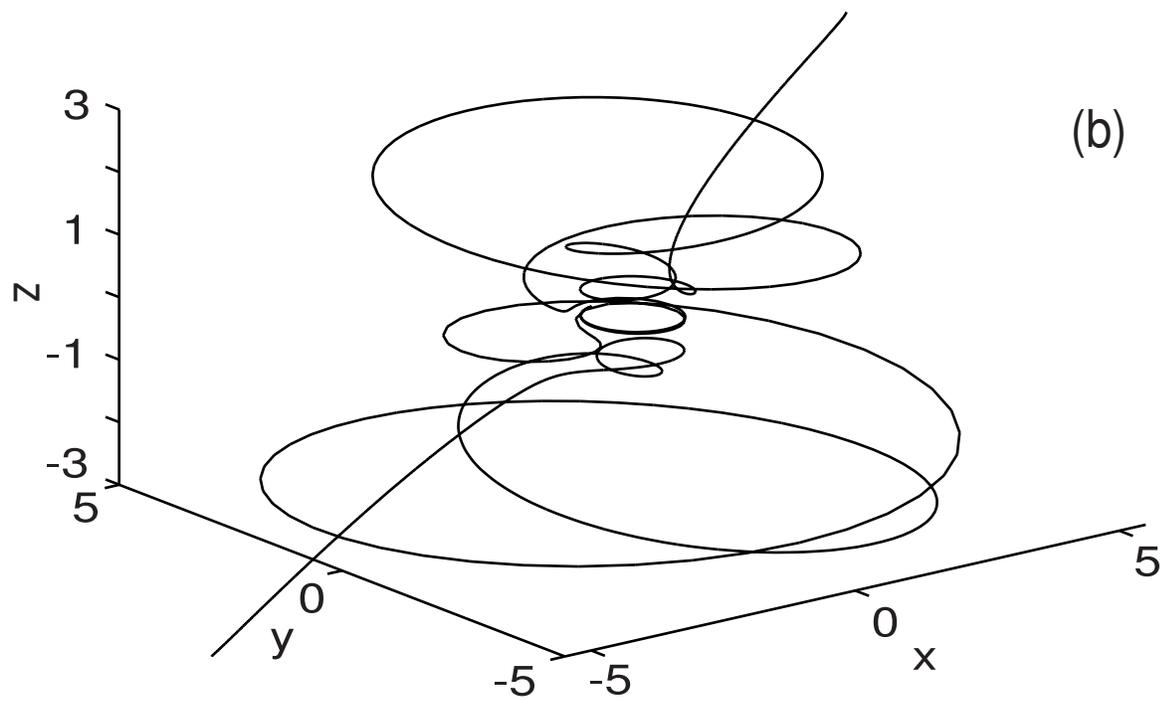
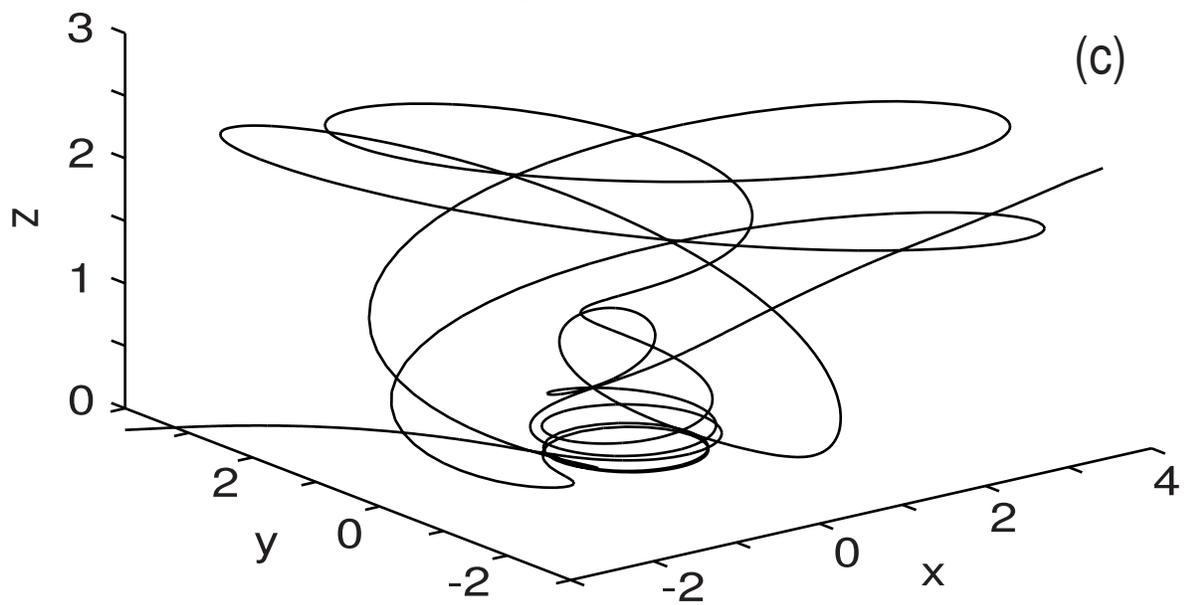
\end{document}